\documentclass{appolb}

\usepackage{graphicx}
\begin{document}
 
\title{Puzzles at High $p_T$ at RHIC and Their Resolution
\thanks{Presented at ISMD2004}
}
\author{Rudolph C. Hwa
\address{Institute of Theoretical Science and Department of
Physics\\ University of Oregon, Eugene, OR 97403-5203, USA}}

\maketitle

\begin{abstract}
Several puzzles about the data at high $p_T$ in heavy-ion collisions are listed. The resolution of them all is given in the framework of parton recombination. More specifically, it is the recombination of the soft and semi-hard shower partons that enhances the region $3<p_T<9$ GeV/c, and gives rise to the large $p/\pi$ ratio in AuAu collisions. The Cronin effect can be explained in terms of final-state interaction for both $\pi$ and $p$. The structure of jets produced in AuAu is different from that in $pp$ collisions. The suppression of $R_{CP}$ in forward production can also be understood by extending the same hadronization scheme at $\eta=0$ to $\eta>0$ without the introduction of any new physics.
\end{abstract}

\vskip0.5cm

In the past two years several features of the high-$p_T$ data obtained by
various experiments at RHIC are puzzling, and may be regarded as anomalies
according to the ``standard model.''  By standard model I mean that which
has been standard in the treatment of hadron production at high $p_T$,
namely:  a hard scattering of partons, followed by a fragmentation process
that leads to the detected hadron.  What I plan to show in this talk are
evidences that all those anomalies can be resolved when the process of hard
parton fragmentation is replaced by the recombination of soft and shower
partons.  The basic reason why the fragmentation model has worked so well
for high-$p_T$ processes in leptonic and hadronic collisions, but poorly for heavy-ion
collisions, is that there is a large body of soft partons in the latter case, but
absent in the former.  The hadronization of those soft partons by
recombination with the semi-hard partons results in a significant
enhancement in the intermediate-$p_T$ region that is missing in the
fragmentation model.

If hard partons fragment in vacuum, whether or not they have lost energy
while in transit in the dense medium, the fragmentation products should be
independent of the medium.  Thus the ratio of produced hadrons, when all
else are the same, should depend only on the ratio of the fragmentation
functions (FF), $D(z)$.  Given a parton, whether a quark or a gluon, its FF
for the production of a proton $D^p(z)$ is much smaller than that for a pion
$D^{\pi}(z)$.  The observed data reveal several anomalies according to that
picture.

\noindent{\it Anomaly 1.} The ratio of proton to pion, $R_{p/\pi}$, in Au+Au
collisions is approximately 1 at $p_T \approx 3$ GeV/c. 

\noindent{\it Anomaly 2.}  The nuclear modification factor, $R_{CP}$, in
d+Au collisions is greater for $p$ than for $\pi$ at $p_T \approx 3$ GeV/c. 

\noindent{\it Anomaly 3.}  The jet structure, i.e., the distribution of particles
associated with a trigger, is different for jets produced in Au+Au collisions
compared to that for p+p collisions.

\noindent{\it Anomaly 4.}  The azimuthal anisotropy parameter $v_2$ is
larger for baryons than for mesons for $p_T \geq 2$ GeV/c. 

\noindent{\it Another irregularity.}  Forward-backward asymmetry.

Time and space do not permit all the topics above to be addressed
adequately.  I give only a sketch here.

How can recombination solve the puzzles?  First of all, let it be understood on
general grounds that when a multi-parton state is to hadronize, it is far more
efficient for a $q$ and $\bar{q}$ to recombine than for a higher momentum
$q$ to fragment, assuming that the parton distribution is falling rapidly in
momentum.  That is simply because recombination involves the addition of
two lower momenta of $q$ and $\bar{q}$, where the densities are higher,
whereas fragmentation involves first the creation of a parton at higher
momentum (at a cost in yield), and then the production of a hadron at some
momentum fraction at the cost of another factor of suppression.  The
comparison is meaningful only when there are many soft partons moving
collinearly with a hard parton, which is the case for heavy-ion collisions, but
not for leptonic and hadronic collisions.

The fragmentation process makes use of a phenomenological FF because it
describes the non-perturbative process of hadronization that cannot be
calculated in pQCD.  Thus $D(z)$ represents a black box, in which there are
gluon radiation, quark pair creation, etc., that generate a shower of partons
before hadronization.  Although the density of shower partons cannot be
calculated from first principles, those partons are nevertheless there, and their
momentum distributions can possibly be determined phenomenologically. 
That is what we have done in the framework of recombination \cite{hy}, in
which we write
\begin{eqnarray}
xD(x) = \int{dx_1 \over x_1}{dx_2 \over x_2}F_{q\bar{q}'}(x_1,
x_2)R_M(x_1, x_2, x),
\label{1}
\end{eqnarray}
where $F_{q\bar{q}'}$ is the joint distribution of a shower quark $q$ and a
shower antiquark $\bar{q}'$ that recombine to form a meson $M$.  $R_M$  is
the corresponding recombination function.  The FF on the LHS is
known from phenomenological analysis of leptonic and hadronic processes. 
$R_M$  is known from previous work on the recombination model.  So
$F_{q\bar{q}'}$ can be determined.  If we assume that $F_{q\bar{q}'}$ has
the factorizable form $S^q_iS^{\bar{q}'}_i$, where $i$ labels the hard parton
that initiates the shower, then there are 5 types of $S^j_i$ to be determined
from 5 types of $D_i$, where $i$ takes on the species:  $u, d, s, g$, and $j$ is
allowed to be $u, d, s$, but not $g$, on the grounds that gluon conversion to
$q\bar{q}$ relieves the burden of considering direct hadronization of
gluons.  The distributions $S^j_i(x)$ of the shower partons have been
determined in \cite{hy} at a fixed $Q = 10$ GeV/c.  The $Q^2$ evolution of 
$S^j_i(x)$ was not considered, although  it constitutes an interesting project
in its own right.  On the basis that hadron production in the intermediate
$p_T$ region at RHIC depends crucially on the recombination of soft and
shower partons, but not sensitively on the virtuality of $S^j_i(x)$ we have
proceeded to the study of the consequences of considering the shower partons
in heavy-ion collisions \cite{hy2}, and found some remarkable results.

For pion production at large $p_T$ the inclusive distribution is 
\begin{eqnarray} 
p{dN_{\pi}  \over  dp} = \int {dp_1 \over p_1}{dp_2 \over p_2}F_{q\bar{q}'}
(p_1, p_2) R_{\pi}(p_1, p_2, p),
\label{2}
\end{eqnarray}
where
\begin{eqnarray}
R_{\pi}(p_1, p_2, p) = {p_1p_2 \over p} \delta (p_1+ p_2 - p).
\label{3}
\end{eqnarray}
Similar equations can be written for proton production \cite{hy2}.  The
essence of recombination is then in what one should include for
$F_{u\bar{d}}$ in case of $\pi^+$, say, and for $F_{uud}$ for $p$.  If we
denote thermal parton distribution by ${\cal T}$ and shower parton
distributions by ${\cal S}$, then they ought to be 
\begin{eqnarray} 
F_{u\bar{d}} = {\cal TT} + {\cal TS} + {\cal SS} ,
\label{4}
\end{eqnarray}
\begin{eqnarray} 
F_{uud} = {\cal TTT} + {\cal TTS} + {\cal TSS} + {\cal SSS} ,
\label{5}
\end{eqnarray}
where the pure ${\cal T}$ terms give the soft component, and the pure ${\cal
S}$ terms recover the fragmentation component.  It is the mixed terms
involving both ${\cal T}$ and ${\cal S}$ that are new and dominate the
intermediate-$p_T$ region, as we shall show below.

To proceed, we need to specify ${\cal T}$ and ${\cal S}$.  For ${\cal T}$ we
do not rely on any low-$p_T$ model, but determine it phenomenologically from the
low-$p_T$ data of pion production, using the ${\cal TT}$ term of (\ref{4}) in 
 (\ref{2}).  In that way we can attribute the enhancement at higher $p_T$
directly to the ${\cal TS}$ term without raising any question on the reliability
of the low-$p_T$ model.  For ${\cal S}$ we shall use the distributions $S^j_i$
determined in \cite{hy}, convoluted with the distribution of hard parton $i$ in
Au+Au collisions, and then sum over $i$.  More specifically, we write
\begin{eqnarray} 
{\cal T}(p_1) = Cp_1e^{-p_1/T},
\label{6}
\end{eqnarray} 
\begin{eqnarray} 
{\cal S}(p_2)= \xi \sum_i \int dk\,kf_i(k)\,S_i^j(p_2/k)\ ,
\label{7}
\end{eqnarray}
where $f_i(k)$ is the hard-parton distribution.  $C$ and $T$ are parameters
to be varied to fit $dN_{\pi}/p_Tdp_T$ for $p_T< 2$ GeV/c.  $\xi$ is
the average fraction of hard partons that can get out of the dense medium to
produce showers.  Without discussing the details that can be found in
\cite{hy2}, let me just show the results.

\begin{figure}[b]
\centerline{
\includegraphics[width=10cm]{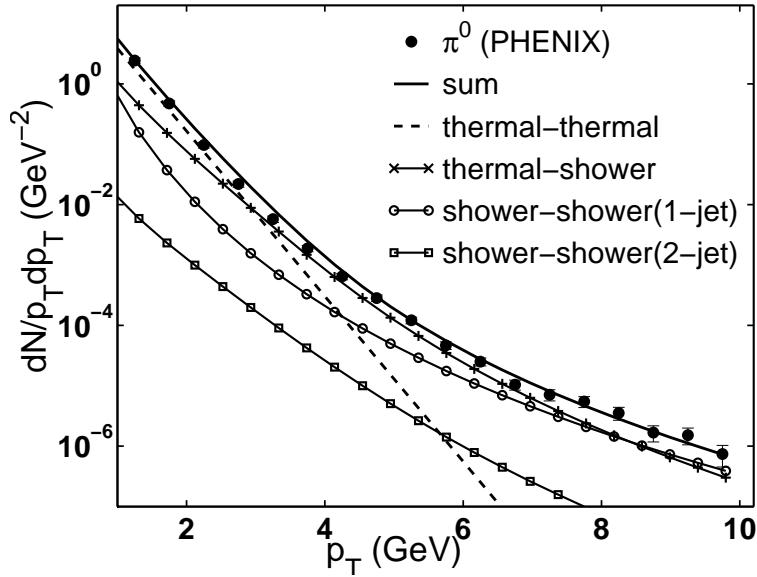}}
\caption{Transverse momentum distribution of $\pi^0$ in Au+Au
collisions. Data are from \cite{ph}.}
\end{figure}

Fig.\ 1 shows the pion distribution that exhibits the dominance of the ${\cal
TS}$ component in the $p_T$ region between 3 and 9 GeV/c.  The sum of all
components agrees well with data \cite{ph}.  Similarly, the ${\cal TTS}$ and
${\cal TSS}$ components dominate over other components of the proton
spectrum in the same $p_T$ region.  The $p/\pi$ ratio is shown in Fig.\ 2;
indeed, it reaches the level of 1 at around $p_T \sim 3$ GeV/c, as observed
\cite{ppi}.  Without the thermal-shower recombination, the proton spectrum
would be too low in the intermediate-$p_T$ region, so the first anomaly in
the fragmentation picture is now satisfactorily resolved.  It should be
mentioned that the large $p/\pi$ ratio has also be obtained in other
recombination/coalescence models using slightly different schemes to
implement the hadronization process \cite{gkl,fmnb}.

\begin{figure}[t]
\centerline{
\includegraphics[width=8cm]{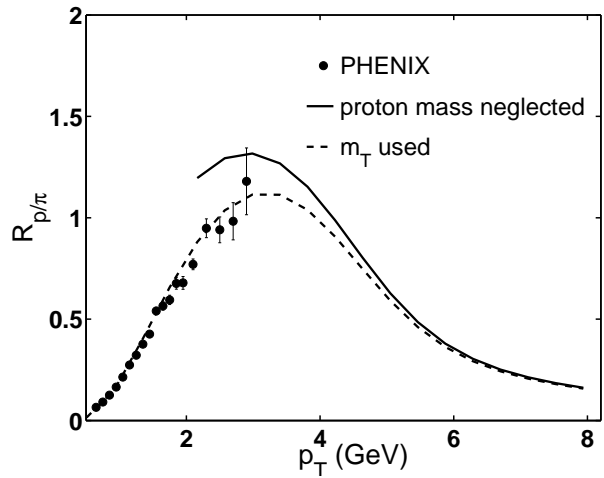}}
\caption{Comparison of calculated $p/\pi$ ratio with data from
\cite{ppi}.}
\end{figure}

The second anomaly concerns the Cronin effect in d+Au collisions.  The
traditional explanation of the effect is that multiple scattering of partons in
the initial state leads to the broadening of the $p_T$ distribution of hadrons
at high $p_T$ in the final state.  If those hadrons are produced by the
fragmentation of high-$p_T$ partons, then $R_{CP}$ for protons should be
much lower than that for pions.  However, the data at RHIC	reveal
$R^p_{CP} >R^{\pi}_{CP }$ for $1 < p_T <3$ GeV/c \cite{fm2}.  This anomaly
can be well explained in the recombination model when the consideration
given above for Au+Au collisions is extended to d+Au collisions
\cite{hy3,hy4}, as shown in Fig.\ 3.  Although the soft partons in d+Au are not thermal in the
sense of Au+Au, the soft component nevertheless plays a similar role and the
recombination of soft and shower partons gives rise to components that can
reproduce the $\pi$ and $p$ spectra in $p_T$ without any initial parton
broadening.  Thus it is the final-state rather than the initial-state interaction
that is mainly responsible for the Cronin effect.  The reason is simply that the
formation of proton by the recombination of 3 quarks, some soft some
semi-hard, is far more effective than the fragmentation of a hard parton.

\begin{figure}[t]
\centerline{
\includegraphics[width=8cm]{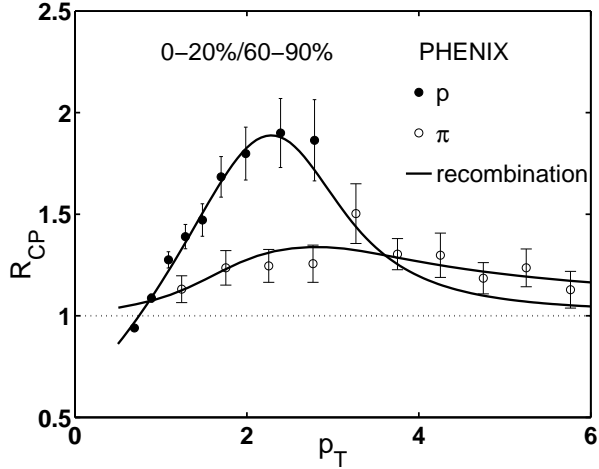}}
\caption{Comparison of calculated  ratios for $R_{CP}$ for $\pi$ and $p$
with data from \cite{fm2}.}
\end{figure}

The third anomaly according to the fragmentation picture is that the jet
structure in Au+Au collisions is different from that in p+p collisions. 
Data from STAR show that the charge multiplicity and total scalar $p_T$ in
the near-side jets are significantly higher for Au+Au than for p+p, even
when the trigger (that is the same for both) is included \cite{fw}.  If the
trigger were not included, one would expect the associated particle
distributions to be drastically different for the two cases.  Such behaviors
cannot follow from the fragmentation of hard partons, once the trigger
momentum is fixed to be the same.  This problem has been studied in the
framework of the recombination model, in which at least two shower partons
in a jet must be considered, one for the trigger, the other for the associated
particle.  Specifically, for $\pi^+$ (trigger) and $\pi^+$ (associated) in a jet the 4-parton distribution has the structure
\begin{eqnarray}
F^{\pi^+\pi^+}_4 =  ({\cal TS}) ({\cal TS})+ ({\cal TS})({\cal SS}) + ({\cal
SS})({\cal TS})\ ,
\label{8}
\end{eqnarray}
where the first pair of parentheses in each term correspond to the trigger, the second  pair the associated particle. We have omitted the term $({\cal SS})({\cal SS})$  in that equation because it is negligible in Au+Au collisions; however, it is the only term that is important in $pp$ collisions. This point clearly reveals the difference between jets produced in heavy-ion and hadronic collisions. The difference becomes even greater when other types of associated particles are included, since the thermal environment in heavy-ion collisions helps the formation of other mesons and baryons in conjunction with shower partons. In Fig.\ 4 we show the distribution associated with a $\pi^+$ trigger when $\pi^+, \pi^-$ and $p$ in the jets are all included \cite{hy5}. The data \cite{fw} are for all charged hadrons in both the trigger and the associated particles, and are therefore not exactly what we have calculated. Nevertheless, the agreement is remarkably good.

\begin{figure}[h]
\centerline{
\includegraphics[width=10cm]{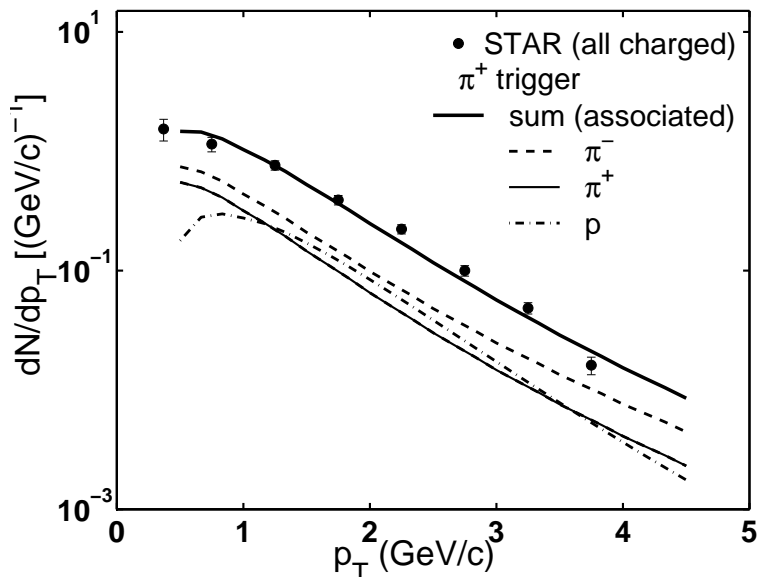}}
\caption{Transverse momentum distribution of $\pi^+, \pi^-$ and $p$ associated with a $\pi^+$ trigger in central Au+Au collisions. Data are from \cite{fw} for all charged hadrons.}
\end{figure}

The fourth anomaly concerns elliptic flow where $v_2$ for baryon exceeds
that for meson.  This phenomenon has nicely been explained by the
coalescence model \cite{mv}, and the scaling of $v_2$ with the number of
constituent quarks remarkably verified by the STAR data \cite{st2}.

The final issue to be mentioned here is about the production of
interme-diate-$p_T$ hadrons at large forward rapidity in d+Au collisions.  It
is the region where high hopes have been raised for the verification of a
signature of color glass condensate.  As with pQCD, the hadronization
mechanism is fragmentation.  BRAHMS data already show that $R_{CP}$ at
$\eta = 3.2$ rises no higher than 0.5 at $p_T \sim 3$ GeV/c \cite{ia}.  This
suppression is regarded as evidence for gluon saturation \cite{jjm}.  However,
before novel physics is invoked, it is reasonable to ask whether the
phenomenon can be understood in the conventional way, i.e., by
extrapolating what is known to work at midrapidity to the forward region.  We
have preliminary results that show a general agreement between the data and
the expectation from parton recombination at all $\eta$ and $p_T$.  The
spectra at forward rapidities are suppressed because there are less soft partons
as $\eta$ is increased, resulting in less hadrons formed that rely on soft
partons recombining with shower partons.  At $\eta = 3.2$ there are so few
hard partons that most hadrons are the result of soft-soft recombination.  
  We have also studied the
backward-forward asymmetry and found that there is no need for a transition
of basic physics from multiple scattering in the initial-state interaction on the
$\eta<0$ side to gluon saturation on the $\eta > 0$ side.

Our emphasis on the hadronization process in the final state provides a
universal framework for the description of particle production at all $\eta$
and $p_T$, at all centralities.  In that framework of interpreting the existing
high-$p_T$ data from RHIC there are no features that are puzzling.  In a sense
that may be disappointing, since exciting new physics usually comes with
anomalies.  However, it is far better to have no puzzles than to be misled by
false anomalies.

All the theoretical results reported here were done in collaboration with C.\ B.\
Yang.  This work was supported, in part, by the U.\ S.\ Department of Energy
under Grant No. DE-FG02-96ER40972.


\end{document}